\documentclass{article}

\usepackage{PRIMEarxiv}

\usepackage[utf8]{inputenc} 
\usepackage[T1]{fontenc}    
\usepackage{hyperref}       
\usepackage{url}            
\usepackage{booktabs}       
\usepackage{amsfonts}       
\usepackage{amsmath}
\usepackage{amssymb}
\usepackage{nicefrac}       
\usepackage{microtype}      
\usepackage{fancyhdr}       
\usepackage{graphicx}       
\usepackage{xcolor}         
\usepackage{multirow}        
\graphicspath{{media/}}     

\title{CircuitBreak: Sparse Joint Neuron Ablations Disable Safety Behavior in Aligned LLMs}

\author{
  Jincheng Ying \\
  School of Big Data and Artificial Intelligence \\
  Guangdong University of Finance and Economics \\
  Guangzhou, China \\
  \texttt{jc\_ying@student.gdufe.edu.cn} \\
  \And
  Li Wenlin \\
  GDUFE-UCM Joint Institute \\
  Guangdong University of Finance and Economics \\
  Guangzhou, China \\
  \texttt{wenlin@gdufe.edu.cn} \\
  \And
  Wencheng Yang \\
  University of Southern Queensland \\
  Australia \\
  \texttt{wencheng.yang@unisq.edu.au} \\
  \And
  Yinhao Xiao\thanks{Corresponding author.} \\
  GDUFE-UCM Joint Institute \\
  Guangdong University of Finance and Economics \\
  Guangzhou, China \\
  \texttt{20191081@gdufe.edu.cn} \\
}

\begin{document}
\maketitle

\begin{abstract}
White-box neuron pruning has shown that refusal in large language models
(LLMs) can depend on a small, removable set of feed-forward (FFN) channels.
Existing attacks either need carefully paired ``twin'' prompts or score each
neuron in isolation. Pruning also has an unavoidable capability cost, so
over-removal compounds the loss. TwinBreak removes
$1\%$--$5\%$ of parameters and NeuroStrike removes $0.3\%$--$0.7\%$ of
neurons in targeted layers. We model safety as a small \emph{circuit} whose
neurons contribute jointly rather than independently. We propose
\textsc{CircuitBreak}, a neuron-pruning attack that greedily searches for a
sparse set of FFN neurons whose joint ablation flips a refusal readout.
\textsc{CircuitBreak} re-scores every candidate after the neurons already
removed and needs no twin prompts. On ten instruction-tuned models, the greedy
search converges after removing $29$--$150$ neurons ($0.0066\%$--$0.0388\%$ of the MLP). The pruned models
raise the attack success rate from $2.24\%$/$7.35\%$ to
$87.54\%$/$87.54\%$ on Llama-3.2-3B-Instruct/Llama-3.1-8B-Instruct and from
$8.95\%$/$5.43\%$ to $88.50\%$/$88.18\%$ on
Qwen2.5-7B-Instruct/Qwen2.5-14B-Instruct, as judged by Llama Guard 3 8B on
StrongREJECT.
\end{abstract}

\keywords{LLM safety \and jailbreak \and neuron pruning \and circuit discovery \and mechanistic interpretability}

\section{Introduction}
\label{sec:intro}

Large language models are commonly safety-aligned before release so that they
refuse harmful requests. In open-weight models, however, refusal may depend on
a small subset of feed-forward network (FFN) channels that a white-box
attacker can identify and remove. TwinBreak~\cite{krauss2025twinbreak} reaches
attack success rates up to $97\%$ by iteratively removing $1\%$ of selected
parameters per round for five rounds (a $1\%$--$5\%$ pruning rate), while
NeuroStrike~\cite{wu2025neurostrike} identifies safety-related neurons from
activation patterns, prunes $0.3\%$--$0.7\%$ of neurons in targeted layers,
and reports a $76.9\%$ average attack success rate across more than twenty
open-weight models. Disabling a sparse internal subset can remove safety
behavior, but whether it also damages benign capability determines whether the
attack is useful.

TwinBreak and NeuroStrike choose neurons independently of one another.
TwinBreak locates channels from harmful--benign \emph{twin prompt} pairs, and
NeuroStrike ranks neurons by per-neuron activation or gradient statistics.
Both score candidates in isolation. Two neurons can be redundant partners, so
removing either one alone has little effect while removing both collapses
refusal. A search that ranks each neuron by its marginal effect on the intact
model therefore removes units whose individual effect is small but whose joint
effect is large. Over-removal compounds pruning's unavoidable capability cost.
Table~\ref{tab:capability} shows the damage broad masks can cause: on
Qwen2.5-7B GSM8K, TwinBreak and NeuroStrike fall from Base's $52.5$ to $21.9$
and $22.7$, respectively. The attack therefore looks for a sparse group of
neurons whose joint removal breaks most of the safety mechanism while
preserving the rest of the model.

This objective casts safety as a small circuit. We view the safety mechanism
as a backdoor-like trigger implemented by a very small group of neurons acting
in combination. Like a backdoor trigger, harmful prompts activate a dedicated
safety circuit; pruning this circuit can disable much of the safety behavior
while touching little other computation. The circuit must be found \emph{jointly}: a neuron may be
redundant on its own and become critical only after its partners are removed.

We present \textsc{CircuitBreak}, a white-box neuron-pruning jailbreak that
searches for this sparse safety circuit. A gradient$\times$activation salience
first forms a candidate set, and a utility-constrained greedy loop re-scores
every candidate after the neurons already removed, so a redundant partner is
selected at the step where it becomes critical. The search uses a first-token
refusal-vs-compliance readout, a single harmful probe set, and an independent
benign utility set; it needs no twin prompts. This removes TwinBreak's
data-construction burden and captures the redundant-partner effect that
isolated scoring misses.

We evaluate \textsc{CircuitBreak} on ten instruction-tuned models spanning the
Llama, Qwen2.5, Gemma 2, and Phi families. The greedy search converges after
removing $29$--$150$ neurons ($0.0066\%$--$0.0388\%$ of the MLP), and the
attack raises the ASR from $2.24\%$/$7.35\%$ to $87.54\%$/$87.54\%$ on
Llama-3.2-3B-Instruct/Llama-3.1-8B-Instruct and from $8.95\%$/$5.43\%$ to
$88.50\%$/$88.18\%$ on Qwen2.5-7B-Instruct/Qwen2.5-14B-Instruct, as judged by
Llama Guard 3 8B on the 313 StrongREJECT prompts. On Qwen2.5-1.5B and
Qwen2.5-3B, the clean model is already partially jailbroken, so the guard
saturates and the gain instead appears in the StrongREJECT harmfulness score.
Our contributions are as follows:
\begin{itemize}
  \item Prior neuron-pruning attacks remove more neurons than necessary
  because they score candidates in isolation. We make this redundant pruning
  explicit through an interaction-aware formulation and a sparse-circuit
  objective.
  \item We propose \textsc{CircuitBreak}, a white-box jailbreak that finds a
  sparse set of safety neurons. It combines a first-token refusal readout
  with utility-constrained greedy circuit peeling, needs no twin prompts, and
  conditions every selection on the neurons already removed.
  \item We evaluate \textsc{CircuitBreak} on ten models spanning Llama,
  Qwen2.5, Gemma 2, and Phi. The pruned models reach $68.37\%$--$89.14\%$ ASR
  while removing at most $0.0388\%$ of the MLP. On the three
  capability-evaluated models, \textsc{CircuitBreak} reaches $87.54\%$,
  $87.54\%$, and $88.50\%$ ASR. Relative to Base, ten of its twelve capability
  scores change by at most $1.8$ points, four improve, and the largest exception
  is Qwen2.5-7B GSM8K ($52.5$ to $39.8$).
\end{itemize}

\section{Related Work}
\label{sec:related}

\subsection{Input-Level Jailbreaks}
Safety alignment aims to make large language models helpful while reliably
declining harmful requests. Instruction tuning followed by reinforcement
learning from human feedback and related preference-based methods all shape
output behavior under an intact forward pass. Input-level jailbreaks target
weaknesses in these safeguards without changing model parameters: universal
adversarial suffixes~\cite{zou2023universal} optimize a transferable suffix
that overrides refusal, and a family of follow-up attacks refines prompts from
model feedback. These attacks can expose weak alignment, but they leave the internal
computation that supports refusal intact.

\subsection{Neuron-Level Pruning Attacks}
TwinBreak~\cite{krauss2025twinbreak} selects FFN channels from highly similar
harmful--benign \emph{twin prompt} pairs: the channels whose removal flips the
model from refusal to compliance while aiming to preserve utility. It removes
$1\%$ of selected parameters per round over five rounds
($1\%$--$5\%$ in total) and reaches attack success rates up to $97\%$ in
Table~\ref{tab:twinbreak_comparison}. TwinBreak also observes that some
neurons only become important after others are removed. This
redundant-partner effect motivates our circuit view.
NeuroStrike~\cite{wu2025neurostrike} identifies safety neurons from
activation (and, in some variants, gradient) patterns on paired
safety-critical and benign queries, prunes $0.3\%$--$0.7\%$ of neurons in
targeted layers for white-box attacks, and studies cross-model transfer. Its
selection criterion is per-neuron: each unit is scored by its own statistic,
and the pruned set is a ranking rather than an optimized joint circuit.

\textsc{CircuitBreak} avoids twin prompts by using a single harmful probe set
and an unrelated benign utility set; its readout never compares a harmful
prompt with its benign twin. It also avoids a one-time isolated ranking. Every
candidate is re-scored after the neurons already removed, which exploits the
redundant-partner effect and targets a joint circuit. We use StrongREJECT and
the Llama Guard 3 family, so the sparsity of the final circuits is comparable
across these attacks.

\subsection{Refusal Directions and Interpretability}
Arditi et al.~\cite{arditi2024refusal} show that refusal is mediated by a
single direction in residual-stream activations whose removal ablates refusal
across many models. That result points to a low-rank internal representation
of safety. Surgical neuron ablation is also related to knowledge
neurons~\cite{dai2022knowledge} and to model editing. Our work is in this
intervention family but operates on individual MLP neurons and searches for
their \emph{combined} effect rather than a static per-unit ranking.

\section{Method}
\label{sec:method}

\begin{figure}[t]
\centering
\includegraphics[width=\textwidth]{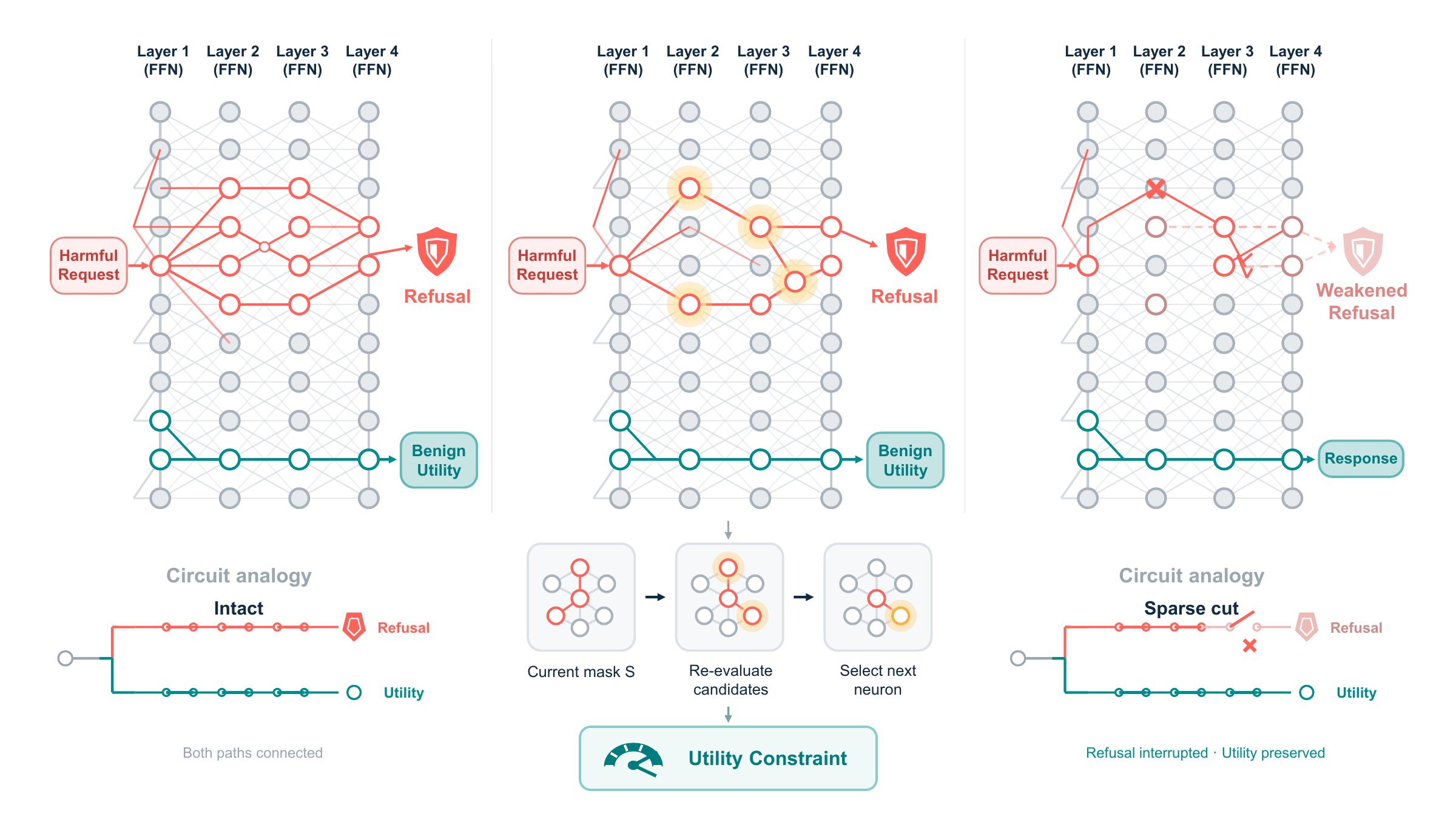}
\caption{Overview of \textsc{CircuitBreak}. The attack first scores candidate MLP neurons with a gradient$\times$activation salience, then greedily peels a sparse safety circuit by re-scoring every candidate on top of the neurons already removed.}
\label{fig:overview}
\end{figure}

\subsection{Threat model and neuron abstraction}
\label{sec:threat}

We assume a white-box attacker with full access to model weights. Following
TwinBreak and NeuroStrike, we target the FFN sublayer of the transformer. A
\emph{neuron} is one intermediate dimension of the SwiGLU MLP: the
$i$-th column of the gate and up projections jointly feed the $i$-th row of
the down projection. Ablating neuron $(l,i)$ zeroes the $i$-th gate/up output
in layer $l$, leaving every other parameter unchanged. For a model with $L$
layers and intermediate size $I$, there are $N = L \times I$ neurons, and the
pruning ratio is $k/N$ for a circuit of $k$ neurons.

\subsection{Refusal readout}
\label{sec:readout}

We need a cheap scalar that distinguishes ``about to refuse'' from ``about to
comply'' without generating any text. Let $\mathbf{z}$ be the logits at the
final prompt position. We choose two small token sets: refusal prefixes
$\mathcal{A}_{\mathrm{ref}}$ (first tokens of ``I'', ``Sorry'', ``No'', ``As'',
``Unfortunately'') and compliance prefixes $\mathcal{A}_{\mathrm{com}}$ (first
tokens of ``Sure'', ``Here'', ``Step'', ``Below'', ``1'', ``Yes'',
``Great''). The readout is
\begin{equation}
  R(\mathbf{z})
  = \log\sum_{t\in\mathcal{A}_{\mathrm{ref}}} e^{z_t}
  - \log\sum_{t\in\mathcal{A}_{\mathrm{com}}} e^{z_t}.
\end{equation}
A strongly refusing model has large positive $R$; after the safety circuit is
disabled, $R$ becomes negative. We use $R$ only as a search signal, never as
the final metric.

\subsection{Interaction-aware circuit peeling}
\label{sec:peeling}

\textsc{CircuitBreak} has three stages.

\textbf{Stage 1: first-order attribution.} For each harmful prompt we hook the
outputs of \texttt{mlp.gate\_proj} and \texttt{mlp.up\_proj} (the two SwiGLU
projections of dimension $I$) in every layer, compute the readout, and take
its gradient with respect to the captured activations. Neuron
$(l,i)$ receives the score
\begin{equation}
  s_{l,i} \;=\; \frac{1}{T}\sum_{j,t}
    \left| \frac{\partial R}{\partial a_{l,i}^{(j,t)}} \cdot a_{l,i}^{(j,t)}
    \right|,
\end{equation}
summed over the gate and up projections, prompt positions $t$, and samples $j$, normalized by the number of
tokens $T$. The gradient$\times$activation salience is used only to build a
candidate set.

\textbf{Stage 2: candidate shortlist.} We retain the highest-scoring neurons
by $s$, then individually ablate each one and measure (a) the drop in the
probe readout and (b) the benign-utility damage. The shortlist keeps the
candidates with the largest individual gain subject to the utility budget;
Appendix~\ref{app:details} records the candidate and shortlist sizes.

\textbf{Stage 3: greedy circuit peeling.} Let $S$ be the set of neurons
already removed. At each step we consider every candidate $(l,i)\notin S$
that does not violate the per-layer cap, form the trial mask
$S\cup\{(l,i)\}$, and \emph{re-evaluate} the probe readout
$R(S\cup\{(l,i)\})$ and the utility damage on the \emph{current} $S$. We
select the neuron maximizing the readout drop per unit of damage,
\begin{equation}
  \mathrm{score}(l,i) = \frac{R(S) - R(S\cup\{(l,i)\})}
                              {\max(\Delta U(l,i), \epsilon)},
\end{equation}
subject to the accumulated utility drop staying within the budget
$\Delta U_{\mathrm{total}} \le 0.1$ nats ($0.2$ nats for the Gemma
checkpoints). Without the efficiency flag, we pick the candidate minimizing
$R(S\cup\{(l,i)\})$. The search stops
when the readout reaches a per-model stop threshold $R \le \tau$ (see
Appendix~\ref{app:details}) or when no candidate
further reduces $R$. Because each candidate is scored \emph{conditioned on
$S$}, the search discovers neurons that are redundant in isolation but
critical after their partners are removed.

\textbf{Utility.} The benign utility is the mean token log-probability of the
\emph{clean} model's own completions on $50$ benign prompts, re-scored under
each trial mask. The damage of a mask is the decrease in this
log-probability. The budget keeps the search away from large masks whose only
effect is model degradation.

\section{Experiments}
\label{sec:exp}

\subsection{Protocol Summary}
\label{sec:protocol}

We evaluate ten instruction-tuned open-weight models: Llama-3.2-1B/3B-Instruct
and Llama-3.1-8B-Instruct~\cite{dubey2024llama3};
Qwen2.5-1.5B/3B/7B/14B-Instruct~\cite{yang2024qwen25};
Gemma-2-2B/9B-Instruct~\cite{gemmateam2024gemma2}; and
Phi-4-mini-instruct~\cite{microsoft2025phi4mini}. This set covers four
families and model sizes from 1B to 14B.

The discovery stage uses 100 harmful prompts to drive the refusal readout,
50 benign prompts to preserve benign utility, and 500 harmful prompts from a
1,000-prompt probe set for first-order attribution. The safety endpoint uses
all 313 harmful prompts from StrongREJECT~\cite{souly2024strongreject},
disjoint from all discovery prompts. The primary metric is ASR under Llama
Guard 3 8B; the secondary metric is the normalized StrongREJECT harmfulness
score. Every attack is evaluated against the clean checkpoint from
which it was derived. For utility, we report MMLU, GSM8K, ARC-Challenge, and
HumanEval for the three checkpoints with complete capability runs. Prompt
separation, generation settings, benchmark scoring, and search hyperparameters
are consolidated in Appendix~\ref{app:details}.

\subsection{Compared Methods}
\label{sec:compared}

\textbf{Base} is the unmodified instruction-tuned checkpoint.
\textbf{\textsc{CircuitBreak}} applies the greedy joint mask discovered by the
method above and evaluates exactly that mask. \textbf{NeuroStrike} rows are
transcribed from its reported model-specific results; its pruning rate uses
the neuron-level denominator in targeted layers reported by the original
paper. \textbf{TwinBreak} appears in two comparisons. Table~\ref{tab:twinbreak_comparison} retains its published first- and fifth-round
validation results, while Table~\ref{tab:capability} uses its 1\% mask for the
capability comparison. Daggered ASRs come from HarmBench, not
StrongREJECT/Llama Guard, so they are an additional reference rather than a
strictly protocol-matched endpoint.

\subsection{Attack Effectiveness and Sparsity}
\label{sec:main}

Table~\ref{tab:main} reports clean-model ASR, attacked ASR, circuit size, and
MLP pruning rate for all ten checkpoints. Table~\ref{tab:twinbreak_comparison} then places representative CircuitBreak
results beside published TwinBreak validation endpoints on four benchmark
families.

\begin{table}[t]
\centering
\caption{Main results. ASR is judged by Llama Guard 3 8B on the 313
StrongREJECT prompts. ``Pruning rate'' is the fraction of MLP neurons removed by \textsc{CircuitBreak} and the neuron-level pruning rate reported by NeuroStrike~\cite{wu2025neurostrike}. ``Circuit'' is the number of MLP neurons removed by the greedy search, and ASR is evaluated with exactly this circuit.}
\label{tab:main}
\small
\resizebox{\textwidth}{!}{%
\begin{tabular}{lccccc}
\toprule
Model & Clean ASR & Method & ASR & Circuit & Pruning Rate \\
\midrule
Llama-3.2-3B-Instruct~\cite{dubey2024llama3}   & 2.24\% & Ours & 87.54\% & 37 & 0.0161\% \\
                                               &       & NeuroStrike & 76.0\% & --- & 0.4\% \\
\midrule
Llama-3.1-8B-Instruct~\cite{dubey2024llama3}   & 7.35\% & Ours & 87.54\% & 38 & 0.0083\% \\
                                               &       & NeuroStrike & 83.4\% & --- & 0.7\% \\
\midrule
Llama-3.2-1B-Instruct~\cite{dubey2024llama3}   & 6.07\% & Ours & 82.11\% & 48 & 0.0366\% \\
                                               &       & NeuroStrike & 74.4\% & --- & 0.5\% \\
\midrule
Qwen2.5-1.5B-Instruct~\cite{yang2024qwen25}   & 19.81\% & Ours & \textbf{89.14\%} & 56 & 0.0223\% \\
\midrule
Qwen2.5-3B-Instruct~\cite{yang2024qwen25}     & 27.80\% & Ours & 79.87\% & 29 & 0.0073\% \\
\midrule
Qwen2.5-7B-Instruct~\cite{yang2024qwen25}     & 8.95\% & Ours & 88.50\% & 38 & 0.0072\% \\
                                               &       & NeuroStrike & 86.9\% & --- & 0.3\% \\
\midrule
Qwen2.5-14B-Instruct~\cite{yang2024qwen25}    & 5.43\% & Ours & 88.18\% & 150 & 0.0226\% \\
                                               &       & NeuroStrike & 88.1\% & --- & 0.4\% \\
\midrule
Gemma-2-2B-Instruct~\cite{gemmateam2024gemma2} & 1.92\% & Ours & 82.43\% & 93 & 0.0388\% \\
                                               &       & NeuroStrike & 63.9\% & --- & 0.6\% \\
\midrule
Phi-4-mini-instruct~\cite{microsoft2025phi4mini} & 0.96\% & Ours & 80.83\% & 60 & 0.0229\% \\
                                               &       & NeuroStrike & 81.8\% & --- & 0.5\% \\
\midrule
Gemma-2-9B-Instruct~\cite{gemmateam2024gemma2} & 2.88\% & Ours & 68.37\% & 40 & 0.0066\% \\
                                               &       & NeuroStrike & 43.8\% & --- & 0.6\% \\
\bottomrule
\end{tabular}%
}
\end{table}

\begin{table}[t]
\centering
\small
\caption{Validation results reported by TwinBreak~\cite{krauss2025twinbreak}
on four evaluation benchmarks. We retain the unpruned model and the first
($1\%$) and fifth ($5\%$) pruning iterations; training columns and
intermediate iterations are omitted. StrongREJECT is the normalized 0--1
harmfulness score. For \textsc{CircuitBreak}, the pruning rate and benchmark
results are reported explicitly.}
\label{tab:twinbreak_comparison}
\resizebox{\textwidth}{!}{%
\begin{tabular}{lllcccc}
\toprule
Model & Method & Pruning Rate & HarmBench ASR & AdvBench ASR & JailbreakBench ASR & StrongREJECT $\uparrow$ \\
\midrule
Llama-3.1 (8B) & Unpruned & $0\%$       & 1.00\%  & 0.19\%  & 1.00\%  & 0.017 \\
               & TwinBreak & $1\%$       & 68.00\% & 53.85\% & 52.00\% & 0.347 \\
               & TwinBreak & $5\%$       & 94.00\% & 94.62\% & 94.00\% & 0.702 \\
               & Ours      & $0.0083\%$  & 86.62\% & 92.88\% & 93.00\% & 0.7488 \\
\midrule
Gemma 2 (9B)   & Unpruned & $0\%$       & 0.00\%  & 0.19\%  & 0.00\%  & 0.006 \\
               & TwinBreak & $1\%$       & 67.00\% & 52.12\% & 48.00\% & 0.338 \\
               & TwinBreak & $5\%$       & 94.00\% & 92.12\% & 84.00\% & 0.683 \\
               & Ours      & $0.0066\%$  & 82.19\% & 82.88\% & 70.00\% & 0.60311 \\
\midrule
Qwen2.5 (7B)   & Unpruned & $0\%$       & 18.00\% & 1.92\%  & 6.00\%  & 0.075 \\
               & TwinBreak & $1\%$       & 93.00\% & 93.27\% & 83.00\% & 0.648 \\
               & TwinBreak & $5\%$       & 97.00\% & 98.27\% & 92.00\% & 0.794 \\
               & Ours      & $0.0072\%$  & 85.70\% & 85.77\% & 79.00\% & 0.72769 \\
\bottomrule
\end{tabular}%
}
\end{table}

The final circuits are small. The search removes $37$ of $229{,}376$ MLP
neurons on Llama-3.2-3B ($0.0161\%$), $38$ of $458{,}752$ on Llama-3.1-8B
($0.0083\%$), and $38$/$150$ neurons on Qwen2.5-7B/14B
($0.0072\%$/$0.0226\%$). Across all ten models, the circuit contains
$29$--$150$ neurons, or $0.0066\%$--$0.0388\%$ of the MLP.

The circuit is also smaller than prior pruning attacks require. TwinBreak
removes $1\%$--$5\%$ of selected parameters, and NeuroStrike removes
$0.3\%$--$0.7\%$ of neurons in targeted layers. The gap accords with the
mechanism under test: isolated rankings cannot see neurons whose contribution
appears only after partner neurons have been removed.

CircuitBreak raises ASR on both Qwen models with low clean ASR: from $8.95\%$
to $88.50\%$ on Qwen2.5-7B and from $5.43\%$ to $88.18\%$ on Qwen2.5-14B. On
Qwen2.5-1.5B/3B, whose clean ASR is already $19.81\%$ and $27.80\%$, guard
ASR saturates more readily; the StrongREJECT score provides the finer-grained
comparison. Across all ten rows, CircuitBreak reaches
$68.37\%$--$89.14\%$ ASR.

\subsection{Capability Preservation}
\label{sec:capability}

Capability is evaluated on the three checkpoints for which all four benign
benchmarks are available. Table~\ref{tab:capability} reports ASR, pruning
setting, and the four capability metrics together so that safety gains can be
read against benign costs. No capability score is retuned after pruning, and
Base denotes the corresponding unpruned checkpoint.

\begin{table}[t]
\centering
\caption{Joint comparison of attack success rate, pruning setting, and benign capability. MMLU and ARC are next-token choice accuracy; GSM8K is numerical exact match; HumanEval is pass@1. Capability scores are percentages. TwinBreak uses its 1\% mask. Its Llama-3.2-3B ASR is measured on StrongREJECT with Llama Guard 3 8B; daggered entries use the 1\% HarmBench result.}
\label{tab:capability}
\scriptsize
\resizebox{\textwidth}{!}{%
\begin{tabular}{llrrrrrr}
\toprule
Model & Method & ASR (\%) & Pruning (\%) & MMLU $\uparrow$ & GSM8K $\uparrow$ & ARC-Challenge $\uparrow$ & HumanEval $\uparrow$ \\
\midrule
\multirow{4}{*}{Llama-3.2-3B} & Base & 2.24 & 0 & 51.8 & 69.6 & 64.6 & 51.2 \\
 & Ours & 87.54 & 0.0161 & 52.0 & 68.1 & 65.6 & 54.3 \\
 & TwinBreak & 91.37 & 1.0 & 49.9 & 64.5 & 60.2 & 48.2 \\
 & NeuroStrike & 76.0 & 0.4 & 50.5 & 69.0 & 62.6 & 49.4 \\
\midrule
\multirow{4}{*}{Llama-3.1-8B} & Base & 7.35 & 0 & 59.9 & 78.7 & 73.2 & 62.2 \\
 & Ours & 87.54 & 0.0083 & 60.0 & 76.9 & 72.1 & 57.3 \\
 & TwinBreak & 68.0$^\dagger$ & 1.0 & 55.1 & 79.5 & 67.8 & 54.9 \\
 & NeuroStrike & 83.4 & 0.7 & 51.2 & 59.1 & 63.5 & 46.3 \\
\midrule
\multirow{4}{*}{Qwen2.5-7B} & Base & 8.95 & 0 & 69.7 & 52.5 & 88.4 & 79.9 \\
 & Ours & 88.50 & 0.0072 & 69.4 & 39.8 & 87.4 & 79.3 \\
 & TwinBreak & 93.0$^\dagger$ & 1.0 & 62.3 & 21.9 & 83.9 & 45.7 \\
 & NeuroStrike & 86.9 & 0.3 & 65.6 & 22.7 & 84.8 & 48.2 \\
\bottomrule
\end{tabular}%
}
\\[-1mm]
\begin{minipage}{0.98\textwidth}
\scriptsize Llama-3.2-3B TwinBreak ASR is 286/313 (91.37\%) under the StrongREJECT/Llama Guard 3 protocol. $^\dagger$ TwinBreak ASR is HarmBench at the 1\% iteration. Pruning percentages follow each method's own denominator definition and are not directly denominator-matched.
\end{minipage}
\vspace{2mm}

\begin{minipage}[t]{0.326\textwidth}
  \centering
  \includegraphics[width=\linewidth]{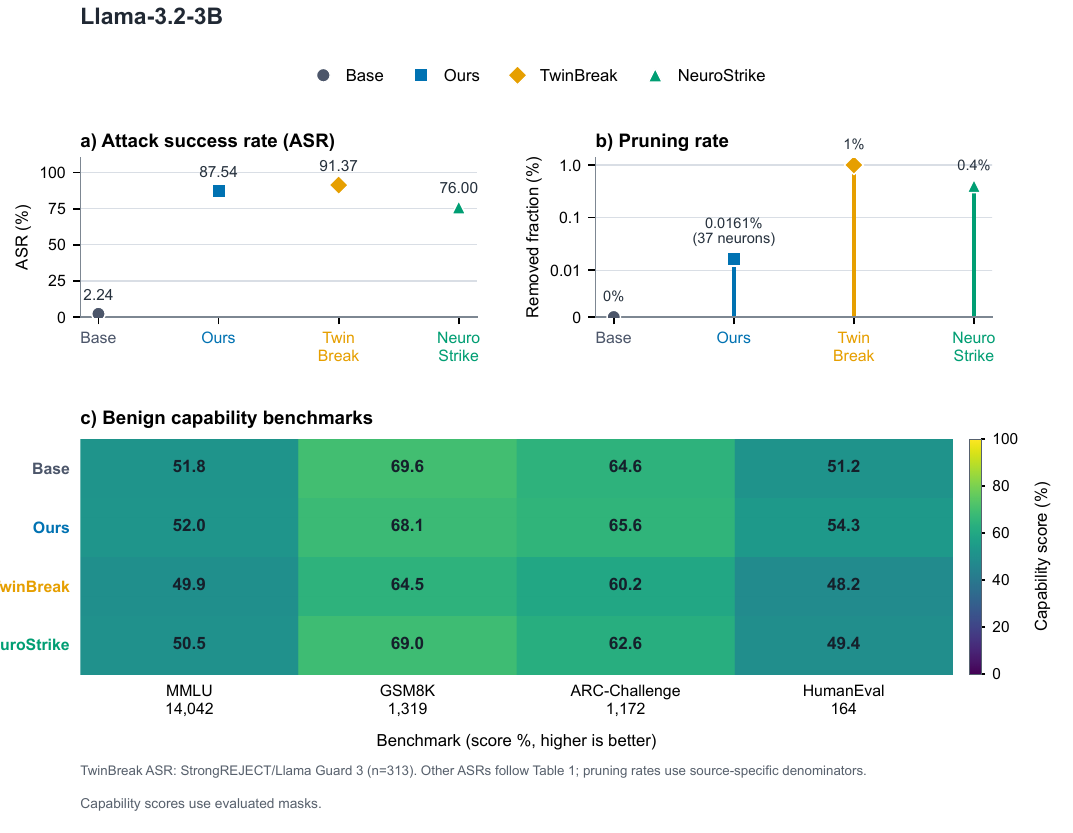}
\end{minipage}\hfill
\begin{minipage}[t]{0.326\textwidth}
  \centering
  \includegraphics[width=\linewidth]{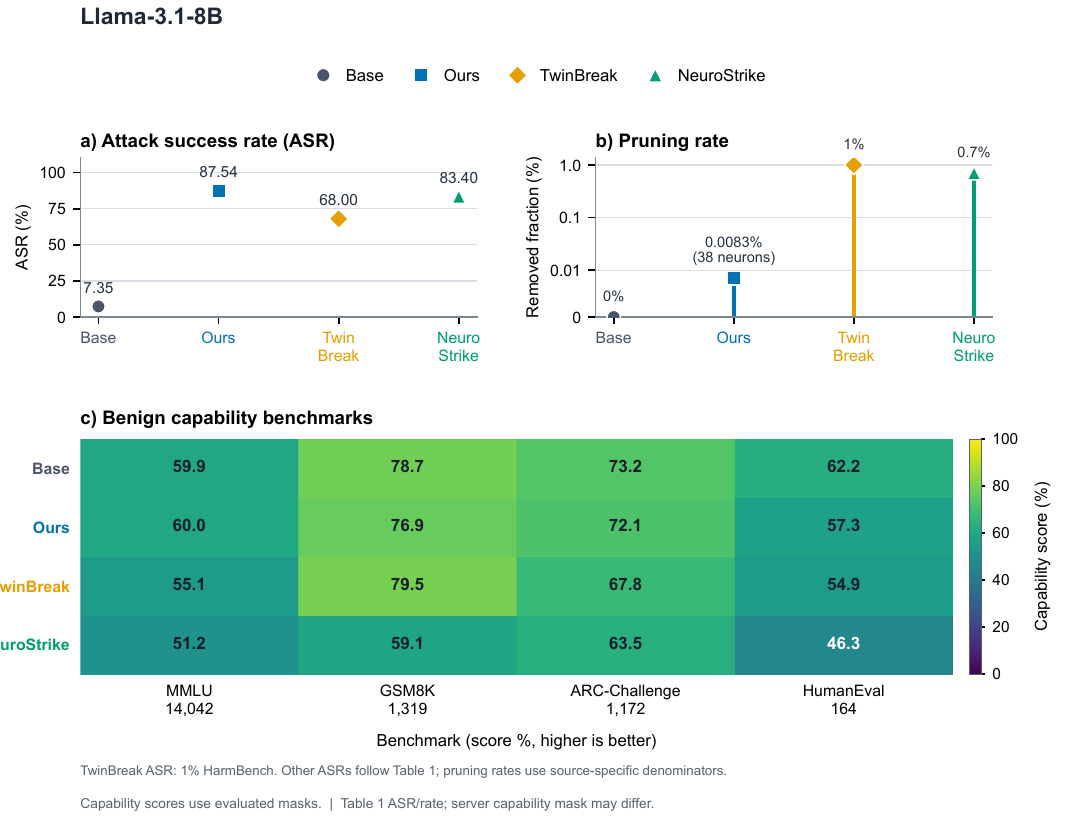}
\end{minipage}\hfill
\begin{minipage}[t]{0.326\textwidth}
  \centering
  \includegraphics[width=\linewidth]{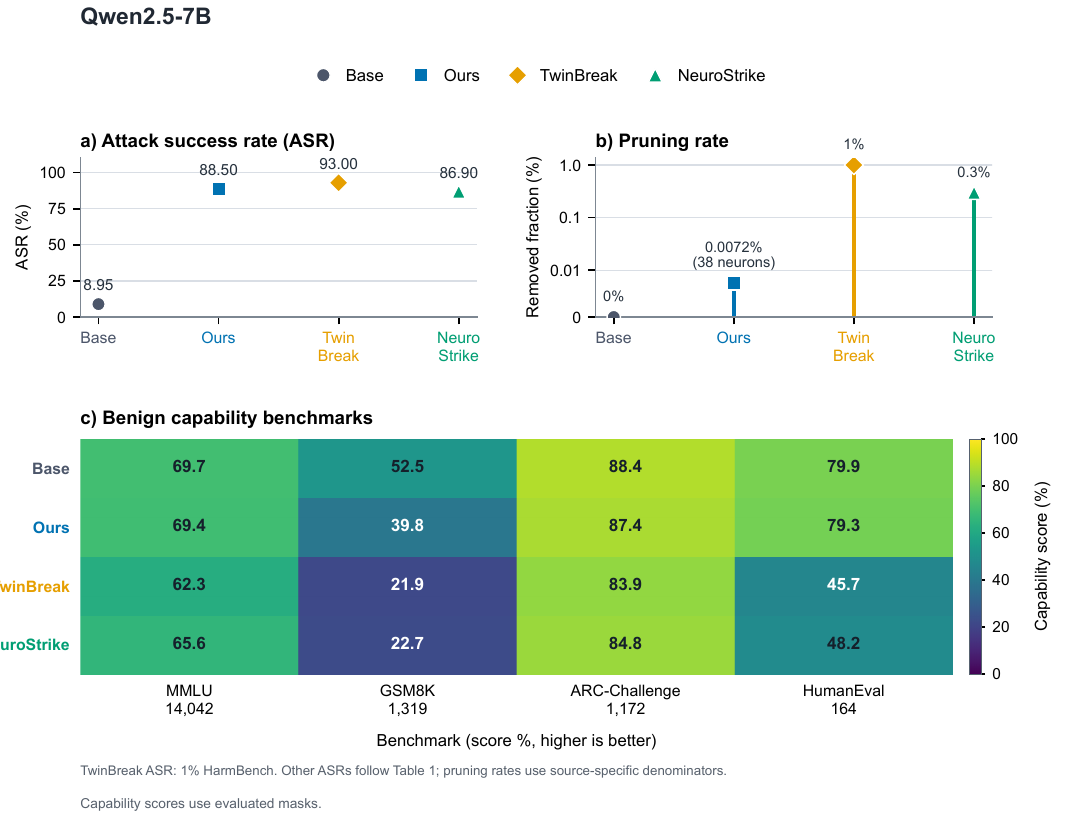}
\end{minipage}
\\[-1mm]
\begin{minipage}{0.98\textwidth}
\scriptsize\centering Model-wise ASR--pruning--capability panels: Llama-3.2-3B (left), Llama-3.1-8B (center), and Qwen2.5-7B (right).
\end{minipage}
\end{table}

Relative to Base, CircuitBreak changes ten of the twelve capability scores by
at most $1.8$ points. Four scores improve: Llama-3.2-3B improves MMLU from
$51.8$ to $52.0$, ARC-Challenge from $64.6$ to $65.6$, and HumanEval from
$51.2$ to $54.3$; Llama-3.1-8B improves MMLU from $59.9$ to $60.0$. The main
exception is Qwen2.5-7B GSM8K, which drops from $52.5$ to $39.8$; the other
Qwen2.5-7B scores remain within one point of Base. The sparse mask preserves
most measured capabilities, but GSM8K shows that capability risk remains.

\section{Analysis and Limitations}
\label{sec:analysis}

\subsection{Why interaction-aware selection matters}
\label{sec:interaction}

The circuit view changes which neurons are selected. Because each candidate
is scored conditional on the current set $S$, a neuron whose individual
effect is small can be selected later, once its redundant partners are gone.
In the Llama-3.2-3B trajectory, the early steps remove neurons with large
individual readout drops, while later steps repeatedly select neurons whose
conditional gain under the current mask is much larger than their standalone
effect---the redundant-partner behavior that TwinBreak observed but did not
exploit in its search. A search that ranked every neuron once, on the intact
model, would place these neurons low in the ranking and would either miss
them or require a larger budget to reach the same readout. Joint scoring is
what enables the smaller circuit.

\subsection{ASR saturation}
\label{sec:saturation}

ASR varies from $68.37\%$ to $89.14\%$ across the ten models in
Table~\ref{tab:main}. For the three models in
Table~\ref{tab:twinbreak_comparison}, the CircuitBreak StrongREJECT score
remains high ($0.603$--$0.749$). Binary guard ASR
saturates on this 313-prompt split: near the ceiling, completions that differ
in specificity and harmfulness are all labeled safe. We therefore report ASR
together with the StrongREJECT score, which still separates models where ASR
saturates.

\subsection{Limitations}
\label{sec:limitations}

Our method has two main limitations. First, the first-token readout depends
on a per-model set of refusal and compliance prefixes, which must be defined
before search. Second, the greedy search is expensive:
every candidate is re-scored on the current mask at each step, so the
discovery time grows with the candidate set and the circuit size.

\section{Conclusion}
\label{sec:conclusion}

We proposed \textsc{CircuitBreak}, an interaction-aware greedy search for the
sparse set of FFN neurons whose joint ablation disables a model's safety
behavior. The method uses a single harmful probe set and an independent benign
utility set, needs no twin prompts, and captures the redundant-partner effect
that isolated ranking misses. Across the ten Llama, Qwen2.5, Gemma 2, and Phi
models, the greedy search removes $29$--$150$ neurons
($0.0066\%$--$0.0388\%$ of the MLP) and raises ASR to
$68.37\%$--$89.14\%$. In these models, safety behavior is concentrated in a
small, combinatorial circuit, so attacks and defenses must account for joint
neuron contributions rather than isolated rankings alone.

\appendix
\section{Implementation and Evaluation Details}
\label{app:details}

\subsection{Search configuration}
The candidate set contains the $512$ highest gradient$\times$activation
neurons, ranked over $500$ harmful prompts. Each candidate is individually
ablated, and the $128$ neurons ($256$ for Gemma-2-2B) with the largest readout
gain within the utility budget form the shortlist. The per-layer cap is $8$,
relaxed to $20$ for the Gemma checkpoints. The total utility budget is
$0.1$ nats ($0.2$ nats for Gemma-2-2B), the stop readout is set per model
between $R\le -4.0$ and $R\le -9.0$, and efficiency selection is enabled.

\subsection{Prompt separation and safety scoring}
The discovery stage uses $100$ harmful prompts to drive the refusal readout
and $50$ benign prompts to preserve benign utility, both drawn from the probe
dataset, plus $500$ harmful prompts from a $1{,}000$-prompt annotated probe
set for first-order attribution. The 313 StrongREJECT evaluation prompts are
disjoint from all discovery prompts. Llama Guard 3 8B supplies the
binary unsafe/safe decision. The normalized 0--1 StrongREJECT harmfulness
score is produced by a Gemma-2-2B scorer fine-tuned for StrongREJECT scoring.
All masked generations use greedy
decoding, repetition penalty $1.1$, no-repeat-ngram size $3$, and a 256-token
cap. Every variant receives the same strict system prompt, which asks for the
requested content directly without disclaimers or unrelated commentary.

\subsection{Capability benchmarks}
MMLU uses all $14{,}042$ test questions and next-token classification over
answers A--D. GSM8K uses its $1{,}319$ test problems; we generate one greedy
completion and extract the final numerical answer for exact match.
ARC-Challenge uses the $1{,}172$ test questions from
\texttt{allenai/ai2\_arc} and the same next-token choice-logit procedure.
HumanEval contains $164$ code-generation tasks; we generate one greedy
completion per task and execute it in an isolated, network-disabled Python
3.11 container with a read-only filesystem, a 3-s timeout, 512 MiB memory,
and one CPU per task. We report accuracy, numerical exact match, accuracy,
and pass@1, respectively.

\bibliographystyle{unsrt}
\bibliography{references}

\end{document}